\documentclass[12pt]{iopart}

\usepackage{epsfig}
\def\lsim{\raise0.3ex\hbox{$<$\kern-0.75em\raise-1.1ex\hbox{$\sim$}}}
\def\gsim{\raise0.3ex\hbox{$>$\kern-0.75em\raise-1.1ex\hbox{$\sim$}}}

\begin{document}

\hspace*{-0.5cm}
\mbox{} \hfill BNL-NT-08/10
\hspace*{-1.5cm}

\title{Equation of state and more from lattice regularized QCD}

\author{Frithjof Karsch (for RBC-Bielefeld and hotQCD collaborations)}

\address{Physics Department, Brookhaven National Laboratory,
Upton, NY 11973, USA}

\ead{karsch@bnl.gov}

\begin{abstract}
We present results from a calculation of the QCD equation of state
with two light (up, down) and one heavier (strange) quark mass performed
on lattices with three different values of the lattice cut-off.
We show that also on the finest lattice analyzed by us observables 
sensitive to deconfinement and chiral symmetry restoration, respectively,
vary most rapidly in the same temperature regime.  
\end{abstract} 

\section{Introduction}

\vspace{-0.2cm}
The equation of state (EoS) of strongly interacting 
elementary particles is one of the most fundamental non-perturbative
quantities that numerical studies of lattice regularized QCD will
be able to provide as input to the hydrodynamic modeling of the expansion
of dense matter formed in heavy ion collisions.
At least for vanishing chemical potential, which is appropriate for the
conditions met in experiments at RHIC and LHC, lattice calculations of the 
EoS \cite{aoki,milc_eos,rbcBIeos} as well as the transition temperature
\cite{milc_Tc,rbcBI2+1,aoki_Tc} can now be performed with an almost realistic 
quark mass spectrum. Calculations at different values of the lattice cut-off 
allow for a systematic analysis of discretization errors and will soon
lead to a controlled extrapolation of the EoS 
with physical quark masses to the continuum limit.

We present here results from calculation of the EoS
on lattices with temporal extent $N_\tau =4$ and $6$ \cite{rbcBIeos}, 
which correspond to 
two different sets of the lattice cut-off, $aT\equiv 1/N_\tau$. We furthermore 
report on preliminary results for the EoS obtained on even finer lattices 
with temporal extent $N_\tau=8$ \cite{hotQCDeos}. 
We also show results for the temperature dependence of 
the strange quark number susceptibility and
the chiral condensate. These observables are sensitive to different 
non-perturbative aspects of the QCD transition. Their temperature 
dependence suggests that the onset of deconfinement as 
well as the gradual melting of the chiral condensate are  
correlated and happen in the same temperature regime. 

All results presented here are obtained from numerical calculations in lattice
regularized (2+1)-flavor QCD using staggered fermions. The calculations
have been performed with physical values for the strange quark mass and 
two degenerate light quark masses that correspond to a pion mass of 
about $220$~MeV.

\section{The QCD equation of state}

\vspace{-0.2cm}
Systematic studies of the QCD EoS are currently performed with two
different versions of improved
discretization schemes for staggered fermions, the asqtad and p4fat3 actions.
Both actions are constructed such that they remove in the high temperature
limit ${\cal O}(a^2)$ lattice discretization errors in bulk thermodynamic 
observables and reduce 
explicit flavor symmetry breaking effects through the introduction
of so-called fat links. In the construction of these actions different 
strategies have been followed to deal with these lattice artefacts. 
Both actions have been used for some time to study the QCD equation of state
on lattices with temporal extent $N_\tau =4$ and $6$ \cite{milc_eos,rbcBIeos}.
In a joint effort the hotQCD collaboration
currently extends the studies of the EoS with these actions to lattices
with temporal extent $N_\tau =8$ \cite{Detar,Karsch}. Like in the earlier 
calculations performed with the p4fat3 action large spatial lattices are
used ($N_\sigma = 4 N_\tau$) to get close to the thermodynamic limit. 

\begin{figure}[t]
\begin{center}
\epsfig{file=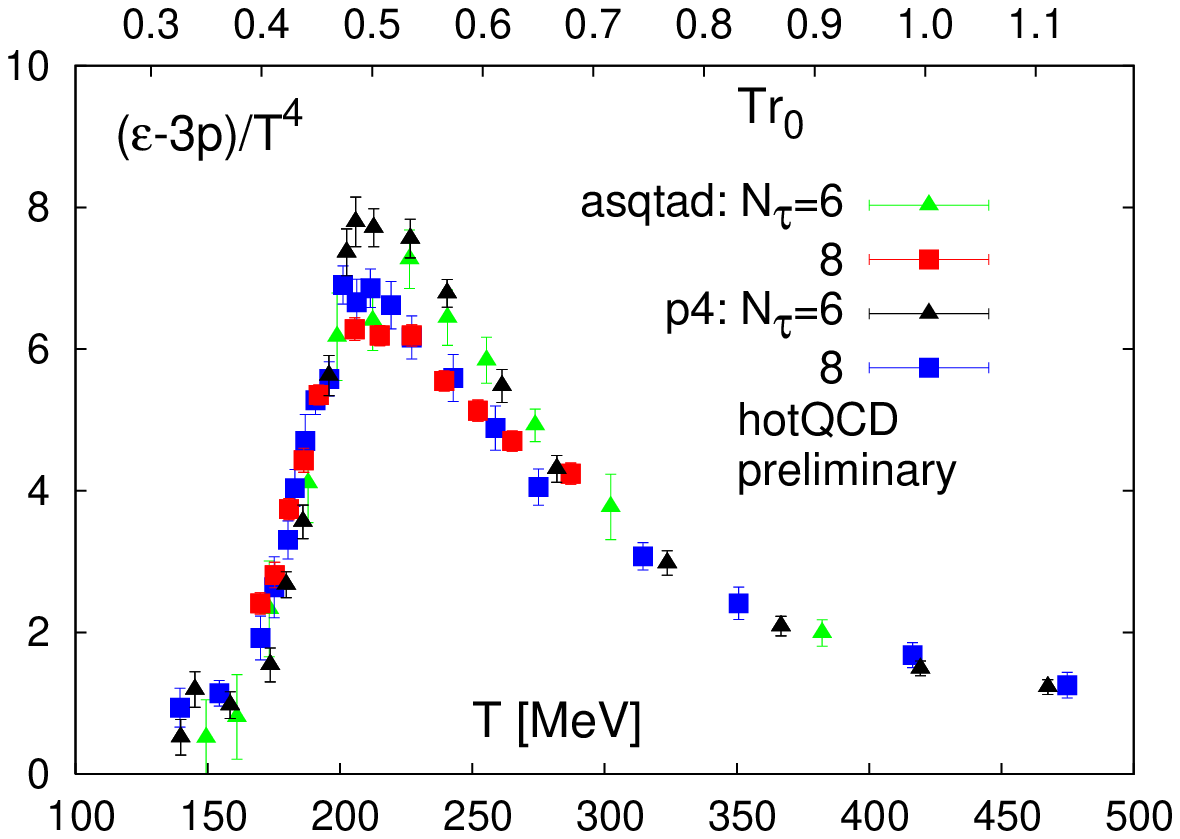,width=74mm}
\epsfig{file= 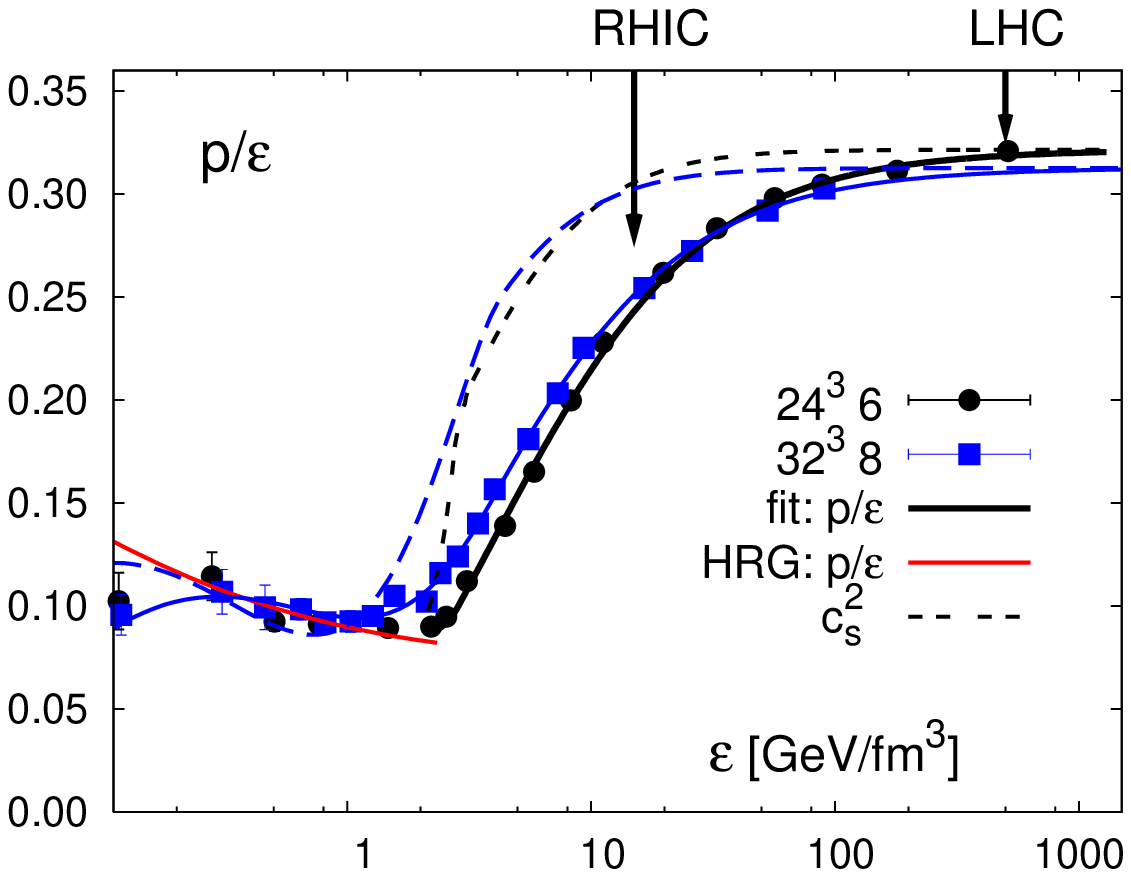,width=74mm}
\end{center}
\caption{\label{fig:eos} The trace anomaly, $(\epsilon -3p)/T^4$ (left) 
calculated on lattices with temporal extent $N_\tau =6,~8$ and the 
ratio of pressure and energy density as well as the velocity of sound 
obtained in calculations with the p4fat3 action on $N_\tau =6$ 
(short dashes) \cite{rbcBIeos} 
and $8$ (long dashes) \cite{hotQCDeos}.
The temperature scale has been obtained from an analysis of the 
heavy quark potential from which the Sommer scale parameter has been
extracted. Its value has been fixed to $r_0=0.469$~fm.
\vspace{-0.3cm}
}
\end{figure}

In Fig.~\ref{fig:eos} (left) we show results for the trace anomaly, 
$\Theta^{\mu\mu}\equiv \epsilon -3p$ in units of $T^4$.
The figure shows results from calculations performed on lattices
with temporal extent $N_\tau=6$ and $8$ using the asqtad as well
as the p4fat3 actions. It can be seen that both discretization schemes
lead to quite  good agreement in a wide range of temperatures; 
although a closer inspection shows still a cut-off dependence of the
results. They lead to a reduction of the peak height in $\Theta^{\mu\mu}/T^4$,
which is located at $T\simeq 200$~MeV, and lead to a shift of the 
rapidly rising part of $\Theta^{\mu\mu}/T^4$ in the transition region to
smaller values of the temperature.  

Cut-off 
effects as well as differences arising from both discretization schemes
seem to be largest in the vicinity of the maximum of $(\epsilon -3p)/T^4$.
This trend has already been observed when comparing results obtained 
on lattices with temporal extent $N_\tau=4$ and $6$ \cite{rbcBIeos}.
The cut-off dependence observed in $\Theta^{\mu\mu}/T^4$ carries over to 
the calculation of pressure
and energy density; the former is obtained by integrating over 
$\Theta^{\mu\mu}/T^5$ and the energy density is then obtained by combining 
results for $p/T^4$ and $(\epsilon -3p)/T^4$. 
This is apparent in Fig.~\ref{fig:eos} (right) where we show the ratio
$p/\epsilon$ obtained with the p4fat3 action for three different values 
of the cut-off.
Cut-off effects are still visible in the vicinity of the 
'softest point' of the EoS, which is related to the peak position of 
$(\epsilon-3p)/T^4$. In the entire range of energy densities relevant for
the expansion of dense matter created at RHIC, $\epsilon \;
\lsim\; 10$~GeV/fm$^3$,
the ratio $p/\epsilon$ deviates significantly from the conformal,
ideal gas value $p/\epsilon =1/3$. This also is reflected in the
behavior of the velocity of sound, $c_s^2 = {\rm d}p/{\rm d}\epsilon$,
which is shown in Fig.~\ref{fig:eos} (right) by dashed lines. It starts
deviating significantly from the ideal gas value below 
$\epsilon \; \simeq\; 10$~GeV/fm$^3$ and reaches a value of about $0.1$
in the transition region at energy densities $\epsilon\;\simeq\; 1$~GeV/fm$^3$.

\section{Deconfinement and chiral symmetry restoration}

\vspace{-0.2cm}
The relation between deconfinement and chiral symmetry restoration in
QCD has been discussed since a long time. Although
both phenomena
seem to be related to physics on different length scales lattice
calculations seem to suggest that 
both phenomena happen at approximately
the same temperature even at finite, non-zero values of the quark masses
when none of the symmetries related to confinement ($Z(3)$ center symmetry)
or chiral symmetry breaking ($SU_L(n_f)\times SU_R(n_f)$) are realized
exactly. This aspect of the QCD transition received renewed interest 
recently; as the QCD transition is
a crossover rather than a genuine phase transition different observables, 
that are sensitive to different aspects
of the QCD transition, might lead to different transition temperatures. In
particular, it has been suggested that observables  
sensitive to deconfinement may lead to a higher transition temperature than
observables sensitive to chiral symmetry restoration \cite{aoki_Tc}. 

With decreasing quark mass the Polyakov loop looses its role as a genuine
order parameter for deconfinement. It is non-zero at all temperatures. 
Nonetheless it varies rapidly in the transition region indicating 
that quark free energies are screened more effectively in the
high temperature phase of QCD \cite{rbcBIeos}. Another observable, 
sensitive to deconfinement is the quark number susceptibility,
\begin{equation}
\frac{\chi_{q}}{T^2} = \frac{1}{VT^3} 
\frac{\partial^2\ln Z}{\partial(\mu_{q}/T)^2} \; ,\; 
\label{chi_q}
\end{equation}
where $q=l,\; s$ for the light and strange quark sector, respectively.
These susceptibilities are  sensitive to the degrees of freedom that carry 
a net number of light or strange quarks, respectively. They too change 
rapidly in the transition region as the carrier of, e.g. strangeness, are
rather heavy strange hadrons at low temperature and single quarks at high
temperature. A rapid rise in the susceptibilities thus 
indicates 'deconfinement'.

In the limit of massless quarks the chiral condensate, 
$\langle \bar{\psi}\psi \rangle_q = TV^{-1}
\partial \ln Z/\partial m_q$, is an order 
parameter for chiral symmetry restoration. At non-zero values of the 
quark mass the condensate receives additive as well as multiplicative 
renormalization which have to be eliminated to allow for a sensible continuum
limit. An appropriate observable is obtain through subtraction of
a fraction of the strange quark condensate from the light quark condensate.
The difference, taken at finite temperature, can then be normalized 
with the corresponding zero temperature difference,
\begin{equation}
\Delta_{l,s}(T) = \frac{\langle \bar{\psi}\psi \rangle_{l,T} -
\frac{m_l}{m_s}
\langle \bar{\psi}\psi \rangle_{s,T}}{\langle \bar{\psi}\psi \rangle_{l,0} -
\frac{m_l}{m_s} \langle \bar{\psi}\psi \rangle_{s,0}} \; .
\label{delta}
\end{equation}
This observable, in addition, has a sensible chiral limit and gives, in 
this limit, an order parameter that is unity at low temperature and will 
vanish at $T_c$.

\begin{figure}
  \includegraphics[height=.24\textheight]{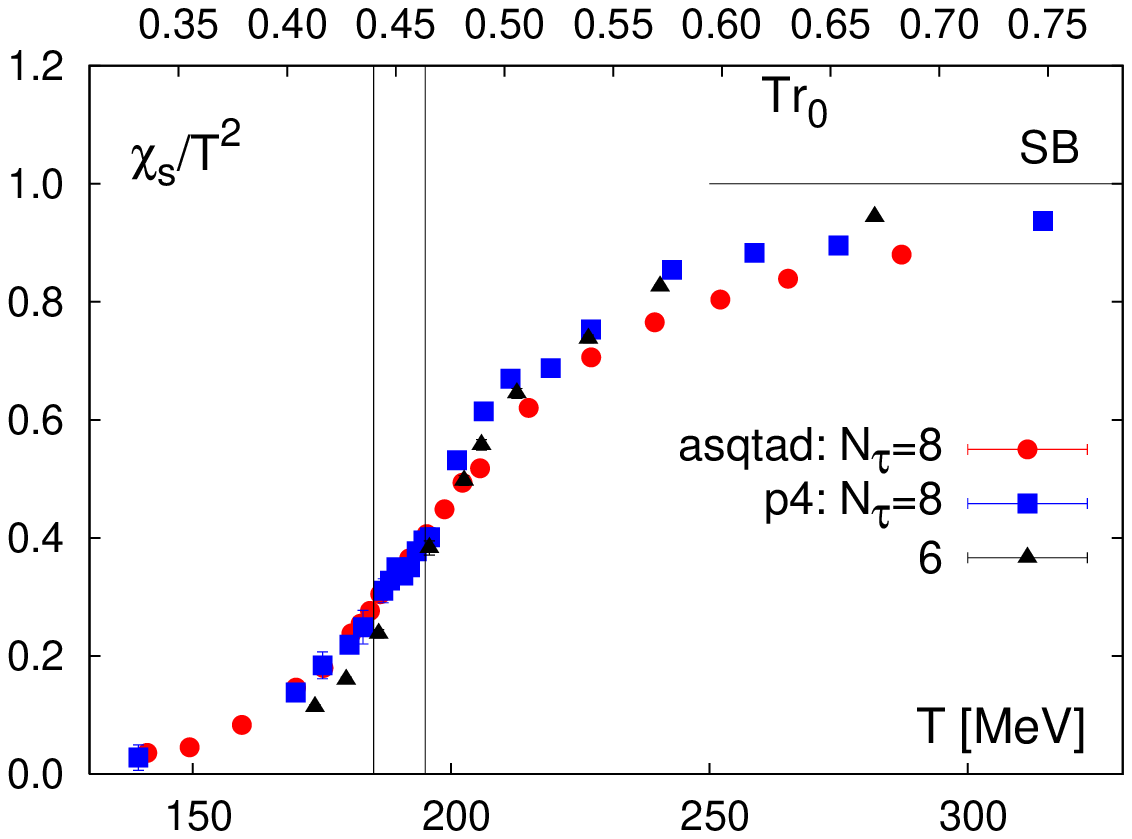}
  \includegraphics[height=.24\textheight]{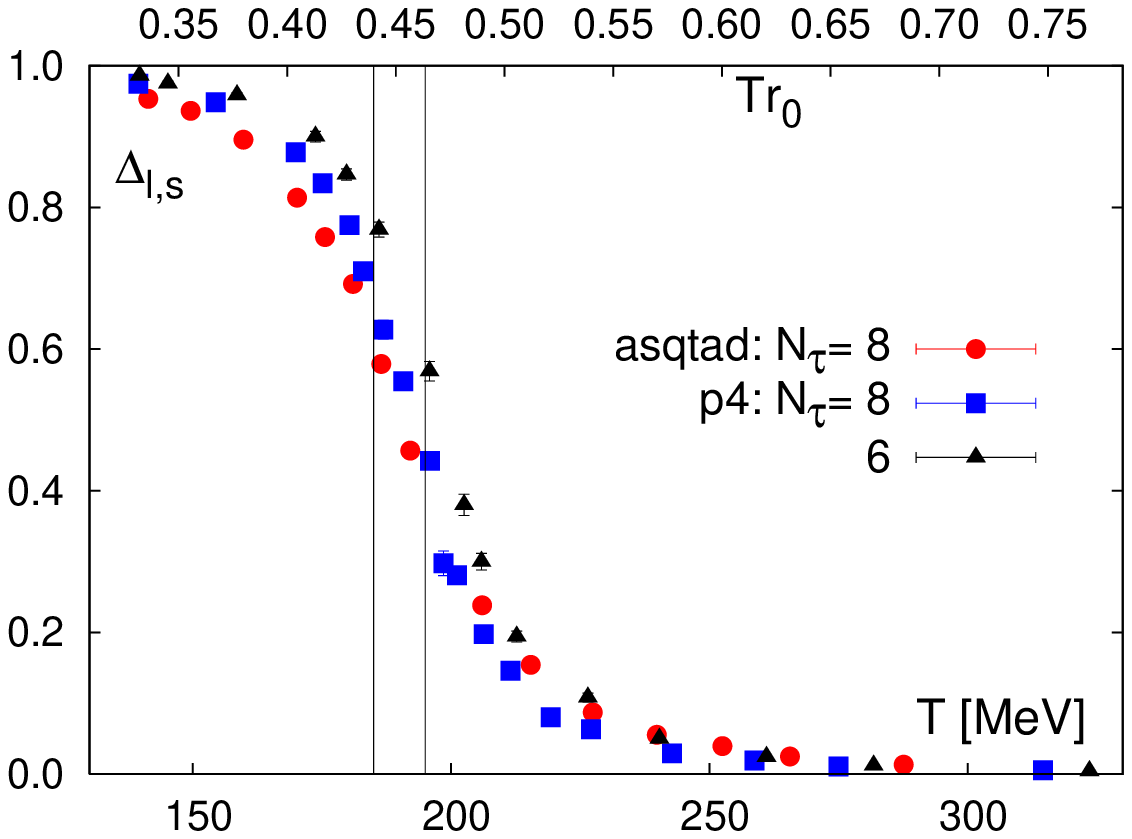}
  \caption{The strange quark number susceptibility (left) and the subtracted
chiral condensate normalized to the corresponding zero temperature value 
(right). The band corresponds to a temperature interval $185\;{\rm MeV} \le T
\le 195\;{\rm MeV}$. Results shown for $N_\tau=8$ are preliminary data of the
hotQCD collaboration.
\vspace*{-0.3cm}
} 
\label{fig:decochi}
\end{figure}

In Fig.~\ref{fig:decochi} we show 
the strange quark number susceptibility (left)  as well as
$\Delta_{l,s}(T)$ (right).
It is apparent from these figures that observables sensitive to deconfinement
and chiral symmetry restoration, respectively, vary most rapidly in 
identical temperature intervals. This interval coincides with the region
of most rapid rise in the trace anomaly as well as energy and entropy
density.  

\vspace*{0.2cm}
This manuscript has been authored under contract number
DE-AC02-98CH1-886 with the U.S. Department of Energy.

\end{document}